\documentclass[12pt]{article}

\usepackage{amssymb}
\usepackage{amsfonts}

\begin{document}

\title{Model of vibrones in quantum photosynthesis \\ as an analog of model of laser}

\author{S.V. Kozyrev}

\maketitle

\centerline{Steklov Mathematical Institute of Russian Academy of Sciences, Moscow, Russia}

\begin{abstract}
Mechanism of vibronic amplification of transport of excitons was discussed in relation to quantum photosynthesis. Vibrones (some modes of vibrations of molecules) are observed experimentally in photosynthetic systems. In the present paper we discuss a model of vibronic amplification of quantum transfer where generation of vibrones as a coherent vibrational mode is described by an analog of semiclassical theory of laser. We consider two models  --- a model of nonequilibrium three level system with vibronic mode, and some variant of a model of laser without inversion. We conjecture that dark states discussed in relation to quantum photosynthesis might be related to mechanism of vibronic ''laser'' without inversion which amplifies the transfer of excitons. We prove that in presence of vibronic mode transfer rate of excitons increases and compute dependence of the transfer rate on parameters of the model.
\end{abstract}

\section{Introduction}

Quantum effects if photosynthesis attract a lot of attention \cite{Engel}, \cite{Scholes}.
In quantum photosynthesis the effect of generation of vibrones is observed. Vibrones are some modes of vibrations of molecules related to transitions between electronic states  (energies of vibrones are equal to Bohr frequencies for corresponding transitions). Relation of vibrones to amplification of transfer of excitons to the photosynthetic reaction center was discussed \cite{1203.5056}, \cite{FerrettiPCCP2013}, \cite{romero2014}, \cite{novoderezhkin2015}, \cite{srep20834}.

In the present text we discuss a model of vibronic amplification of quantum transport of excitons in quantum photosynthesis based on the idea that vibrones can be described as a phononic analog of the laser mode. For description of vibrones we apply the semiclassical theory of laser.

We discuss a variant of semiclassical theory of laser where equations of state for media are described by quantum dissipative dynamics in the Lindblad form. We use the dynamics given by the quantum stochastic limit method  \cite{AcLuVo}, \cite{Notes}. To complete the model we add to Lindblad equations for media the equation for laser mode from the semiclassical laser theory. This approach is a particular case of the standard semiclassical theory of lasers \cite{Haken}, \cite{Scully}.

We show that generation of coherent vibronic mode indeed increases the transfer rate. For the laser mechanism population inversion is necessary, moreover the inversion should exceed some threshold. There exist models of lasers without inversion where so called quantum dark states of the system interacting with laser mode are applied \cite{Koch}, \cite{Harris}, \cite{Mompart}, \cite{Scully}.

Dark states attract attention in quantum optics and in applications to quantum informatics \cite{Dark}. Quantum dark states were investigated experimentally in photosynthetic systems \cite{FerrettiPCCP2013}, \cite{novoderezhkin2015}, \cite{srep20834}. Here we consider a modification of the laser model of vibrones where quantum dark states are generated and vibrones are described by a model of laser without inversion.

In \cite{ArVoKo}, \cite{VoKo}, \cite{KMTV}, \cite{Cosine}, \cite{DarkStates} nonequilibrium quantum dissipative dynamics of the stochastic limit method for quantum theory was applied to modeling of quantum photosynthesis.
Manipulations by quantum states and application to quantum computations and quantum control were considered in \cite{VoKo},  \cite{DarkStates}, \cite{PechenIlyin}, \cite{PechenTru}.
In \cite{OhyaVolovich} application of quantum methods to computations and biology was discussed. Relation of ''global'' and ''local'' approaches in theory of open quantum systems was investigated in \cite{TruVo}.

Exposition of the present paper is as follows. In section 2 we discuss a variant of semiclassical theory of laser for three level system interacting with nonequilibrium environment.
Transfer rate in stationary state and condition of generation of the coherent ''laser'' mode are computed, we show that presence of the coherent vibronic mode amplifes the transfer rate. In section 3 we consider analogous model for four level system which is an analog of ''laser without inversion'' which uses quantum dark states. We show that in this model it is easier to satisfy the condition of generation of the coherent mode (vibrones). Section 4 is the conclusion of the paper.

\section{Laser mechanism for vibrones}

We consider quantum dissipative dynamics of a system with three energy levels $\varepsilon_0<\varepsilon_1<\varepsilon_2$ and Hamiltonian
\begin{equation}\label{H_S}
H_S=\varepsilon_0 |0\rangle\langle 0|+ \varepsilon_1 |1\rangle\langle 1|+\varepsilon_2|2\rangle\langle 2|.
\end{equation}
The physical meaning is as follows --- the system describes excitons in quantum photosynthesis in one exciton approximation, the lower state $|0\rangle$ is the state without exctions, the state $|1\rangle$ describes exciton in the reaction center, the state $|2\rangle$ describes exciton on chromophore. Here $\{|0\rangle,|1\rangle,|2\rangle\}$ is the orthonormal basis in the system Hilbert space ${\mathcal H}_S$.

The system interacts with nonequilibrium environment given by three reservoirs (Bose fields in temperature states with different temperatures). Interaction of the system with light leads to creation of excitons, interaction of excitons with phonons (vibrations of molecules) is related to transfer of excitons to the reaction center, absorption of excitons in the reaction center is described by the additional field of sink. The total Hamiltonian of the system and environment has the form
\begin{equation}\label{H_total}
H=H_S+H_{\rm em}+H_{\rm ph}+H_{\rm sink}+ \lambda\left( H_{I,{\rm em}}+H_{I,{\rm ph}}+H_{I,{\rm sink}}\right).
\end{equation}

Here $H_R$, $R={\rm em},\, {\rm ph},\, {\rm sink}$ are Hamiltonians of the reservoirs (Bose fields)
\begin{equation}\label{H_R}
H_R=\int_{\mathbb{R}^3} \omega_R(k)a^{*}_R(k) a_R(k) dk,
\end{equation}
which act in Bose Hilbert spaces of the reservoirs ${\mathcal H}_R$. The total Hamiltonian of the system and environment acts in the tensor product of the spaces of the system and reservoirs
${\mathcal H}_S\otimes {\mathcal H}_{{\rm em}}\otimes {\mathcal H}_{{\rm ph}}\otimes{\mathcal H}_{{\rm sink}}$.

Each of reservoirs is in the temperature state --- Gaussian mean zero state with quadratic correlation function
$$
\langle a^{*}_{R}(k)a_R(k') \rangle=N_R(k)\delta(k-k'),
$$
where $N_R(k)$ has the form
\begin{equation}\label{temperature}
N_R(k)={1\over{e^{\beta_R\omega_R(k)}-1}},
\end{equation}
and $\beta_R$ is the inverse temperature of the reservoir.

Interaction Hamiltonians $H_{I,{\rm em}}$, $H_{I,{\rm ph}}$, $H_{I,{\rm sink}}$ have forms (\ref{H_em}), (\ref{H_ph}), (\ref{H_sink}) correspondingly. Here $g_{R}(k)$ are form factors of interaction of the system with reservoirs (complex valued functions).

Interaction of the system with light is described by the interaction Hamiltonian
\begin{equation}\label{H_em}
H_{I,{\rm em}} = A_{\rm em}|2\rangle\langle 0| + A^{*}_{\rm em}|0\rangle\langle 2|,\qquad A^{*}_{\rm em}=\int_{\mathbb{R}^3} g_{\rm em}(k)a^{*}_{\rm em}(k)  dk.
\end{equation}

Transport of excitons to the reaction center is related to interaction with phonons
\begin{equation}\label{H_ph}
H_{I,{\rm ph}} = A_{\rm ph}|2\rangle\langle 1| + A^{*}_{\rm ph}|1\rangle\langle 2|,\qquad A^{*}_{\rm ph}=\int_{\mathbb{R}^3} g_{\rm ph}(k)a^{*}_{\rm ph}(k)  dk.
\end{equation}

Absorption of excitons in the reaction center is described by interaction with the sink reservoir
\begin{equation}\label{H_sink}
H_{I,{\rm sink}} = A_{\rm sink}|1\rangle\langle 0| + A^{*}_{\rm sink}|0\rangle\langle 1|,\qquad A^{*}_{\rm sink}=\int_{\mathbb{R}^3} g_{\rm sink}(k)a^{*}_{\rm sink}(k)  dk.
\end{equation}

For the model under investigation dynamics of reduced density matrix of the system interacting with three reservoirs will be generated by the sum of four generators
\begin{equation}\label{total_gen}
\frac{d}{dt}\rho(t)=\left(n_{\rm em}\theta_{\rm em}+n_{\rm ph}\theta_{\rm ph}+i[\cdot, H_{\rm eff}]+n_{\rm sink}\theta_{\rm sink}\right)(\rho(t)),
\end{equation}
where $\theta_R$, $R={\rm em},\, {\rm ph},\, {\rm sink}$ are quantum dissipative generators in the Lindblad form and $n_R$ is the total number of quanta for reservoir $R$.

Contribution $i[\cdot, H_{\rm eff}]$ with effective Hamiltonian
$$
H_{\rm eff}=s(|2\rangle\langle 1|+|1\rangle\langle 2|),\quad s\in \mathbb{R}
$$
will describe vibronic mechanism (analog of interaction with the laser mode). Here $s$  is the amplitude of vibronic mode. Generators $\theta_R$ are obtained using the procedure of the stochastic limit of quantum theory \cite{AcLuVo} (Lamb shift is ignored) and have the following form.

Generation of excitons is described by the photonic generator, $R={\rm em}$, Bohr frequency equals $\omega_{\rm em}=\varepsilon_2-\varepsilon_0$, the temperature is $\beta_{\rm ph}^{-1}=6000K$
\begin{equation}\label{theta_em}
\theta_{\rm em}(\rho)=
2\gamma^{-}_{\rm em}
\left(
\langle 2|\rho|2\rangle |0\rangle\langle 0|
-{1\over 2}
\{\rho,|2\rangle\langle 2|\}\right)+
2\gamma^{+}_{\rm em}
\left(\langle 0| \rho |0\rangle |2\rangle\langle 2|
-{1\over 2}
\{\rho,|0\rangle\langle 0| \}\right).
\end{equation}

Transport of excitons is described by the phononic generator, $R={\rm ph}$, Bohr frequency equals $\omega_{\rm ph}=\varepsilon_2-\varepsilon_1$, the temperature is $\beta_{\rm ph}^{-1}=300K$
\begin{equation}\label{theta_ph}
\theta_{\rm ph}(\rho)=
2\gamma^{-}_{\rm ph}
\left(
\langle 2|\rho|2\rangle |1\rangle\langle 1|
-{1\over 2}
\{\rho,|2\rangle\langle 2|\}\right)+
2\gamma^{+}_{\rm ph}
\left(\langle 1| \rho |1\rangle |2\rangle\langle 2|
-{1\over 2}
\{\rho,|1\rangle\langle 1| \}\right).
\end{equation}

Absorption of excitons is described by the sink generator, $R={\rm sink}$, Bohr frequency equals $\omega_{\rm sink}=\varepsilon_1-\varepsilon_0$, the temperature is $\beta_{\rm sink}^{-1}=300K$
\begin{equation}\label{theta_sink}
\theta_{\rm sink}(\rho)=
2\gamma^{-}_{\rm sink}
\left(
\langle 1|\rho|1\rangle |0\rangle\langle 0|
-{1\over 2}
\{\rho,|1\rangle\langle 1|\}\right)
+2\gamma^{+}_{\rm sink}
\left(
\langle 0|\rho|0\rangle |1\rangle\langle 1|
-{1\over 2}
\{\rho,|0\rangle\langle 0|\}\right).
\end{equation}

Coefficients $\gamma^{\pm}_{R}$ have the form (with $N_R(k)$ given by (\ref{temperature}))
\begin{equation}\label{gamma+}
\gamma^{+}_{R}=\pi\int
|g_{R}(k)|^2\delta(\omega_{R}(k)-\omega_R)N_{R}(k)dk,
\end{equation}
\begin{equation}\label{gamma-}
\gamma^{-}_{R}=\pi\int
|g_{R}(k)|^2\delta(\omega_{R}(k)-\omega_R)(N_{R}(k)+1)dk.
\end{equation}

Value $\gamma^{+}_{R}$ is the rate of induced transitions between corresponding energy levels (pair of levels with energy difference $\omega_R$) due to interaction with the reservoir $R$, value $\gamma^{-}_{R}$ is the sum of rates of spontaneous and induced transitions.

Rates of transitions (for the temperature state of the reservoir) satisfy
$$
\frac{\gamma^{+}_{R}}{\gamma^{-}_{R}}=e^{-\beta_R\omega_R},
$$
where $\omega_R$ is the Bohr frequency of the transition and $\beta_R$ is the inverse temperature of the reservoir, i.e.
$$
\frac{\gamma_{\rm em}^{+}}{\gamma_{\rm em}^{-}}=e^{-\beta_{\rm em}(\varepsilon_2-\varepsilon_0)},\qquad
\frac{\gamma_{\rm ph}^{+}}{\gamma_{\rm ph}^{-}}=e^{-\beta_{\rm ph}(\varepsilon_2-\varepsilon_1)},\qquad
\frac{\gamma_{\rm sink}^{+}}{\gamma_{\rm sink}^{-}}=e^{-\beta_{\rm sink}(\varepsilon_1-\varepsilon_0)}.
$$

We will consider dynamics of the system, described by equation (\ref{total_gen}), in the space of density matrices of the form (including diagonal matrix elements and one pair of off-diagonal elements)
\begin{equation}\label{rho}
\rho=\rho_{22}|2\rangle\langle 2| + \rho_{11}|1\rangle\langle 1| + \rho_{00}|0\rangle\langle 0| + \rho_{21}|2\rangle\langle 1|+ \rho_{12}|1\rangle\langle 2|.
\end{equation}
Generator of the dynamics has the form
\begin{equation}\label{L}
L=n_{\rm em}\theta_{\rm em}+n_{\rm ph}\theta_{\rm ph}+i[\cdot, H_{\rm eff}]+n_{\rm sink}\theta_{\rm sink}.
\end{equation}

Other off-diagonal elements of density matrix will decay exponentially with this dynamics (i.e. we will have decoherence), elements $\rho_{21}$, $\rho_{12}$ can be generated due to interaction with vibronic mode (contribution $i[\cdot, H_{\rm eff}]$ in the generator) with non-zero $s$.

For description of generation of vibronic mode by transitions between energy levels $|2\rangle$ and $|1\rangle$ we use the following equation of semiclassical theory of laser \cite{Haken}
\begin{equation}\label{dsdt1}
\frac{d}{dt}s=-\varkappa s -i \rho_{21}.
\end{equation}
Here $\varkappa$ is the rate of dissipation of vibronic mode.

We will investigate the stationary solution for the dynamics $L(\rho)=0$ and amplitude of vibronic mode $s$ in the stationary state (using (\ref{dsdt1})). We will compute the transition rate of excitons to sink and will prove that in presence of non-zero vibronic mode the rate of transport of excitons increases. We propose to consider this effect as a model of vibronic mechanism of amplification of exciton transport.

\medskip

\noindent{\bf Stationary nonequilibrium set} $L(\rho)=0$ without vibrones ($s=0$) has the form
$$
\rho_{21}=\rho_{12}=0,
$$
\begin{equation}\label{rho22}
\rho_{22}=\frac{
n_{\rm em}n_{\rm ph}\gamma_{\rm em}^{+}\gamma_{\rm ph}^{+}+
n_{\rm em}n_{\rm sink}\gamma_{\rm em}^{+}\gamma_{\rm sink}^{-}+
n_{\rm ph}n_{\rm sink}\gamma_{\rm ph}^{+}\gamma_{\rm sink}^{+}}{\Delta},
\end{equation}
\begin{equation}\label{rho11}
\rho_{11}=\frac{
n_{\rm em}n_{\rm ph}\gamma_{\rm em}^{+}\gamma_{\rm ph}^{-}+
n_{\rm em}n_{\rm sink}\gamma_{\rm em}^{-}\gamma_{\rm sink}^{+}+
n_{\rm ph}n_{\rm sink}\gamma_{\rm ph}^{-}\gamma_{\rm sink}^{+}
}{\Delta},
\end{equation}
\begin{equation}\label{rho00}
\rho_{00}=\frac{
n_{\rm em}n_{\rm ph}\gamma_{\rm em}^{-}\gamma_{\rm ph}^{+}+
n_{\rm em}n_{\rm sink}\gamma_{\rm em}^{-}\gamma_{\rm sink}^{-}+
n_{\rm ph}n_{\rm sink}\gamma_{\rm ph}^{-}\gamma_{\rm sink}^{-}
}{\Delta},
\end{equation}
where the normalization $\Delta$ ensures that the trace of density matrix is equal to one
\begin{equation}\label{Delta}
\Delta=
n_{\rm em}n_{\rm ph}\left(\gamma_{\rm ph}^{+}\gamma_{\rm em}^{+}+
\gamma_{\rm ph}^{-}\gamma_{\rm em}^{+}+
\gamma_{\rm ph}^{+}\gamma_{\rm em}^{-}\right)+
$$ $$+
n_{\rm ph}n_{\rm sink}\left(\gamma_{\rm ph}^{+}\gamma_{\rm sink}^{+}+
\gamma_{\rm ph}^{-}\gamma_{\rm sink}^{+}+
\gamma_{\rm ph}^{-}\gamma_{\rm sink}^{-}\right)+
$$ $$+
n_{\rm em}n_{\rm sink}\left(\gamma_{\rm em}^{+}\gamma_{\rm sink}^{-}+
\gamma_{\rm em}^{-}\gamma_{\rm sink}^{+}+
\gamma_{\rm em}^{-}\gamma_{\rm sink}^{-}\right).
\end{equation}

\medskip

\noindent{\bf Proof.}\quad
Computation of the stationary state $L\rho=0$ in absence of vibrones $s=0$ can be performed as follows. In this case off-diagonal matrix elements $\rho_{21}$, $ \rho_{12}$ in the stationary set will be equal to zero (since for $s=0$ these matrix elements will be eigenvectors of matrix $L$ with negative eigenvalues).

In the diagonal subspace (of dimension three) matrix $L$ will be degenerate (this is degenerate square matrix with sum of lines equal to zero). Therefore the solution of the system of linear equations $L\rho=0$ can be found if we take matrix elements $\rho_{22}$, $\rho_{11}$, $\rho_{00}$ to be equal to minors with alternating signs\footnote{Minor corresponding to matrix element of square matrix is the determinant of matrix obtained from the initial matrix  by eliminating of the line and the column which intersect at the matrix element.} of the first line of the matrix $L$ (in basis $\rho_{22}$, $\rho_{11}$, $\rho_{00}$).
Equation corresponding to the first line of the matrix will correspond to expansion of the determinant of the matrix (which equals zero) along the first line and will be satisfied identically.

Let us consider the second equation of the system $L\rho=0$ and substitute in this equation the same set of variables. Since the matrix $L$ is degenerate the second line of the matrix is a linear combination (actually, a sum with opposite sign) of other lines. Let us subtract from the second line elements of the linear combination corresponding to lines of the matrix except for the first line until only contribution from the first line remains. These operations do not change the determinant (equal to zero). Thus the second equation will be satisfied for the same set of variables, analogous statement holds for other equations. Therefore the set of minors with alternating signs for the first line of the matrix $L$ gives the solution of the system.

If the obtained solution is not equal to zero identically and the matrix $L$ has corank one (which holds for generic case), we will obtain a general form of the stationary state in absence of vibrones $s=0$. The normalization condition $\rho_{22}+\rho_{11}+\rho_{00}=1$ gives the solution (\ref{rho22}), (\ref{rho11}), (\ref{rho00}), (\ref{Delta}) for the stationary state.

\medskip

{\bf Transfer rate of excitons to the sink} equals  (using expression for the sink generator)
$$
F=2n_{\rm sink}\left(\gamma^{-}_{\rm sink}\rho_{11}-\gamma^{+}_{\rm sink}\rho_{00}\right).
$$

For the above stationary state (\ref{rho22}), (\ref{rho11}), (\ref{rho00}), (\ref{Delta}) we get for the flow of excitons to the sink (taking in account $\beta_{\rm ph}=\beta_{\rm sink}$)
\begin{equation}\label{Flux}
F=\frac{2n_{\rm em}n_{\rm ph}n_{\rm sink}\left(\gamma_{\rm em}^{+}\gamma_{\rm ph}^{-}\gamma^{-}_{\rm sink}
-\gamma_{\rm em}^{-}\gamma_{\rm ph}^{+}\gamma^{+}_{\rm sink}\right)
}{\Delta}=
$$ $$
=\frac{2n_{\rm em}n_{\rm ph}n_{\rm sink}\gamma_{\rm em}^{-}\gamma_{\rm ph}^{+}\gamma^{+}_{\rm sink}
}{\Delta}\left(e^{\left(\beta_{\rm ph}-\beta_{\rm em}\right)(\varepsilon_2-\varepsilon_0)}-1\right).
\end{equation}

Dependence of the transfer rate $F$  on values $n_R$ possesses saturating behavior: for any reservoir $R$ for small number of quanta the flow is proportional to $n_R$ and for $n_R\to
\infty$ the flow saturates i.e. tends to constant.

\medskip

\noindent{\bf Case of presence of vibrones}.\quad
For the stationary state in presence of vibrones equation $L\rho=0$ implies
$$
\rho_{21}=-\frac{is}{\mu}(\rho_{22}-\rho_{11}),\quad
\rho_{12}=\frac{is}{\mu}(\rho_{22}-\rho_{11}),\quad \mu=-n_{\rm ph}\left(\gamma^{-}_{\rm ph}+\gamma^{+}_{\rm ph}\right).
$$
i.e. off-diagonal matrix elements are proportional to the inversion $\rho_{22}-\rho_{11}$.
Substitution of these expressions to the system of equations for the stationary state implies
$$
\pmatrix{
-2n_{\rm em}\gamma^{-}_{\rm em}-2n_{\rm ph}\gamma^{-}_{\rm ph}+2\frac{s^2}{\mu};&  2n_{\rm ph}\gamma^{+}_{\rm ph}-2\frac{s^2}{\mu};& 2n_{\rm en}\gamma^{+}_{\rm em}\cr
2n_{\rm ph}\gamma^{-}_{\rm ph}-2\frac{s^2}{\mu}; & -2n_{\rm ph}\gamma^{+}_{\rm ph}-2n_{\rm sink}\gamma^{-}_{\rm sink}+2\frac{s^2}{\mu};& 2n_{\rm sink}\gamma^{+}_{\rm sink}\cr
2n_{\rm en}\gamma^{-}_{\rm em};& 2n_{\rm sink}\gamma^{-}_{\rm sink};& -2n_{\rm en}\gamma^{+}_{\rm em}-2n_{\rm sink}\gamma^{+}_{\rm sink}\cr
}
\pmatrix{\rho_{22}\cr \rho_{11}\cr \rho_{00}\cr }=0.
$$

We get the following recipe for description of the stationary state: the diagonal matrix elements of the stationary density matrix have the form
(\ref{rho22}), (\ref{rho11}), (\ref{rho00}), (\ref{Delta}) where the following transformations were made
\begin{equation}\label{change1}
n_{\rm ph}\gamma_{\rm ph}^{-} \mapsto n_{\rm ph}\gamma_{\rm ph}^{-}-\frac{s^2}{\mu},
\end{equation}
\begin{equation}\label{change2}
n_{\rm ph}\gamma_{\rm ph}^{+}\mapsto n_{\rm ph}\gamma_{\rm ph}^{+} -\frac{s^2}{\mu}.
\end{equation}

Value $\mu$ is negative i.e. for non-zero $s$ this transformation increases the parameters $n_{\rm ph}\gamma_{\rm ph}^{\pm}$.
Equations (\ref{rho22}), (\ref{rho11}) imply for the inversion
\begin{equation}\label{inv3level}
\rho_{22}-\rho_{11}=\frac{
\left(n_{\rm em}\gamma_{\rm em}^{+}+n_{\rm sink}\gamma_{\rm sink}^{+}\right)n_{\rm ph}\left(\gamma_{\rm ph}^{+}-\gamma_{\rm ph}^{-}\right)+
n_{\rm em}n_{\rm sink}\left(\gamma_{\rm em}^{+}\gamma_{\rm sink}^{-}-\gamma_{\rm em}^{-}\gamma_{\rm sink}^{+}\right)
}{\Delta(s)},
\end{equation}
where $\Delta(s)$ is obtained from (\ref{Delta}) by substitutions (\ref{change1}), (\ref{change2}), i.e.
\begin{equation}\label{Delta'}
\Delta(s)=\Delta-\frac{s^2}{\mu}
\left(
n_{\rm em}\left(2\gamma_{\rm em}^{+}+\gamma_{\rm em}^{-}\right)+
n_{\rm sink}\left(2\gamma_{\rm sink}^{+}+\gamma_{\rm sink}^{-}\right)\right)
\end{equation}
and the inversion decreases with $s$.

Expression (\ref{inv3level}) for the inversion in the regime of irreversible absorption of excitons in the reaction center ($\gamma_{\rm sink}^{+}=0$) takes the form
\begin{equation}\label{inv3level1}
\rho_{22}-\rho_{11}=\frac{
n_{\rm em}\gamma_{\rm em}^{+}\left(n_{\rm sink}\gamma_{\rm sink}^{-}-n_{\rm ph}\left(\gamma_{\rm ph}^{-}-\gamma_{\rm ph}^{+}\right)\right)}{\Delta(s)},
\end{equation}
i.e. the inversion is positive if the rate $n_{\rm sink}\gamma_{\rm sink}^{-}$ of absorption of excitons in the reaction center is larger than the rate of transfer of excitons to the reaction center due to spontaneous transitions $n_{\rm ph}\left(\gamma_{\rm ph}^{-}-\gamma_{\rm ph}^{+}\right)$.

Flow of excitons to the sink in presence of vibrones takes the form
\begin{equation}\label{Flux1}
F=\frac{2n_{\rm em}n_{\rm ph}n_{\rm sink}\left(\gamma_{\rm em}^{+}\gamma_{\rm ph}^{-}\gamma^{-}_{\rm sink}
-\gamma_{\rm em}^{-}\gamma_{\rm ph}^{+}\gamma^{+}_{\rm sink}\right)
}{\Delta(s)}
-\frac{s^2}{\mu}
\frac{2n_{\rm em}n_{\rm sink}\left(\gamma_{\rm em}^{+}\gamma^{-}_{\rm sink}
-\gamma_{\rm em}^{-}\gamma^{+}_{\rm sink}\right)
}{\Delta(s)},
\end{equation}
i.e. the flow of excitons increases in presence of vibrones. We propose to consider this effect as a description of vibronic mechanism of amplification of exciton transfer rate.

The laser generation has threshold behavior i.e. in the model under consideration vibronic mechanism of transport amplification is switched on with sufficiently large inversion.
Let us consider equation of generation of vibrones (\ref{dsdt1}) in the stationary state
$$
\varkappa s=-i\rho_{21}=-\frac{s}{\mu}(\rho_{22}-\rho_{11}).
$$

For all values of the parameters there exists the solution $s=0$ (without vibronic mode). Necessary condition of existence of solution with non-zero $s$ --- positivity of inversion $\rho_{22}-\rho_{11}$ (since $\mu$ is negative). For the solution with non-zero vibrones we get for the square of amplitude of the vibronic mode
\begin{equation}\label{s^2}
s^2=\frac{\left(n_{\rm em}\gamma_{\rm em}^{+}+n_{\rm sink}\gamma_{\rm sink}^{+}\right)n_{\rm ph}\left(\gamma_{\rm ph}^{+}-\gamma_{\rm ph}^{-}\right)+
n_{\rm em}n_{\rm sink}\left(\gamma_{\rm em}^{+}\gamma_{\rm sink}^{-}-\gamma_{\rm em}^{-}\gamma_{\rm sink}^{+}\right)-
\Delta n_{\rm ph}\left(\gamma^{-}_{\rm ph}+\gamma^{+}_{\rm ph}\right)\varkappa}
{\varkappa
\left(
n_{\rm em}\left(2\gamma_{\rm em}^{+}+\gamma_{\rm em}^{-}\right)+
n_{\rm sink}\left(2\gamma_{\rm sink}^{+}+\gamma_{\rm sink}^{-}\right)\right)}.
\end{equation}
Here $\Delta$ is given by (\ref{Delta}) (i.e. corresponds to the case without vibrones). Positivity of the above expression is the sufficient condition (in our model) for existence of the vibronic mode.

Positivity of (\ref{s^2}) can be satisfied only for sufficiently small $\varkappa$ (when the numerator is positive), moreover the smaller $\varkappa$ (dissipation of vibrones) will be the larger $s$ (value of vibronic mode) we will obtain in the stationary state. As we discussed above if $s$ increases the flow of excitons $F$ will increase and it will tend to some constant value determined by the level of light and the rate of absorption of excitons in the reaction center.

In approximation of irreversible absorption of excitons in the reaction center ($\gamma_{\rm sink}^{+}=0$) we get for the square of amplitude of the vibronic mode (\ref{s^2})
$$
s^2=\frac{n_{\rm em}\gamma_{\rm em}^{+}\left(n_{\rm sink}\gamma_{\rm sink}^{-}- n_{\rm ph}\left(\gamma_{\rm ph}^{-}-\gamma_{\rm ph}^{+}\right)\right)-
\Delta n_{\rm ph}\left(\gamma^{-}_{\rm ph}+\gamma^{+}_{\rm ph}\right)\varkappa}
{\varkappa
\left(
n_{\rm em}\left(2\gamma_{\rm em}^{+}+\gamma_{\rm em}^{-}\right)+n_{\rm sink}\gamma_{\rm sink}^{-}\right)}
$$
and for the flow of excitons to the sink (\ref{Flux1})
$$
F=\frac{2n_{\rm em}n_{\rm sink}\gamma_{\rm em}^{+}\gamma^{-}_{\rm sink}
}{\Delta(s)}\left(n_{\rm ph}\gamma_{\rm ph}^{-}-\frac{s^2}{\mu}\right),
$$
where $s^2\ge 0$ and $\mu<0$.

Hence {\it in presence of vibrones the transfer rate of excitons to the sink increases (due to presence of positive contribution $- \frac{s^2}{\mu}$ in $n_{\rm ph}\gamma^{-}_{\rm ph}- \frac{s^2}{\mu}$).}

\section{Vibronic laser without inversion}

For the described in the previous section model it is difficult to generate sufficiently large inversion to overcome the laser (vibronic) generation threshold (\ref{s^2}) which is possible only for small dissipation $\varkappa$. For small inversion the vibronic transport amplification mechanism is switched off. To solve this problem it is possible to use the known mechanism of inversion free laser \cite{Koch}, \cite{Harris}, \cite{Mompart}, \cite{Scully}. This kind of lasers use so called quantum dark states. For inversion free laser it is sufficient to create inversion in the subspace where quantum transport operates (not in the total space of density matrices which includes dark states).

Let us discuss the following modification of the introduced in the previous section model. In this model the second (middle) energy level will be almost degenerate. In particular this level could correspond to two interacting molecules (let us note that the photosynthetic reaction center contains the special pair of chlorophylls). In this case we get the energy level splitting --- instead of the state $|1\rangle$ the middle energy level will contain two states $|1\rangle$ and $|1'\rangle$, and the following symmetric and antisymmetric states will be eigenvectors of the system Hamiltonian
$$
|S\rangle=\frac{1}{\sqrt{2}}\left(|1\rangle+|1'\rangle\right),\qquad |A\rangle=\frac{1}{\sqrt{2}}\left(|1\rangle-|1'\rangle\right).
$$

The system Hamiltonian takes the form
\begin{equation}\label{H_S1}
H_S=\varepsilon_0 |0\rangle\langle 0|+ \varepsilon_S |S\rangle\langle S|+ \varepsilon_A |A\rangle\langle A|+\varepsilon_2|2\rangle\langle 2|,\qquad
\varepsilon_0<\varepsilon_S<\varepsilon_A<\varepsilon_2.
\end{equation}

We take for simplicity that the energy difference between $\varepsilon_S$ and $\varepsilon_A$ is small with respect to other Bohr frequencies in the system.

We assume that (using some additional interactions) there exists exchange of population between the symmetric and antisymmetric states. This process can be described using the additional generator of the system density matrix dynamics
\begin{equation}\label{theta_ex}
\theta_{\rm ex}(\rho)=
2\gamma_{\rm ex}
\left(
\langle A|\rho|A\rangle |S\rangle\langle S|
-{1\over 2}
\{\rho,|A\rangle\langle A|\}\right)+
2\gamma_{\rm ex}
\left(
\langle S|\rho|S\rangle |A\rangle\langle A|
-{1\over 2}
\{\rho,|S\rangle\langle S|\}\right).
\end{equation}
Here for simplicity we take the coefficients $\gamma_{\rm ex}$ for two terms in the generator equal.

Thus the total generator of the dynamics of density matrix of the system will have the form
\begin{equation}\label{L1}
L=n_{\rm em}\theta_{\rm em}+n_{\rm ph}\theta_{\rm ph}+i[\cdot, H_{\rm eff}] + n_{\rm ex}\theta_{\rm ex}+n_{\rm sink}\theta_{\rm sink}.
\end{equation}
Here generator $\theta_{\rm em}$ has the same form (\ref{theta_em}) as in the previous section and generators $\theta_{\rm ph}$, $i[\cdot, H_{\rm eff}]$ and $\theta_{\rm sink}$ will be modified as follows.

The sink generator (\ref{theta_sink}) (which describes absorption of excitons in the reaction center) will be substituted by the sum of two terms which instead of $|1\rangle$ will contain $|S\rangle$ and $|A\rangle$ correspondingly (i.e. excitons are absorbed in both states $|S\rangle$ and $|A\rangle$ and values $\gamma^{\pm}_{\rm sink}$ for the symmetric and antisymmetric states are taken equal)
\begin{equation}\label{theta_sink1}
\theta_{\rm sink}(\rho)=
2\gamma^{-}_{\rm sink}
\left(
\langle S|\rho|S\rangle |0\rangle\langle 0|
-{1\over 2}
\{\rho,|S\rangle\langle S|\}\right)
+2\gamma^{+}_{\rm sink}
\left(
\langle 0|\rho|0\rangle |S\rangle\langle S|
-{1\over 2}
\{\rho,|0\rangle\langle 0|\}\right)+
$$
$$
+2\gamma^{-}_{\rm sink}
\left(
\langle A|\rho|A\rangle |0\rangle\langle 0|
-{1\over 2}
\{\rho,|A\rangle\langle A|\}\right)
+2\gamma^{+}_{\rm sink}
\left(
\langle 0|\rho|0\rangle |A\rangle\langle A|
-{1\over 2}
\{\rho,|0\rangle\langle 0|\}\right).
\end{equation}

Generator (\ref{theta_ph}) will be substituted by the generator of the form
\begin{equation}\label{theta_ph1}
\theta_{\rm ph}(\rho)=
2\gamma^{-}_{\rm ph}
\left(
\langle 2|\rho|2\rangle |S\rangle\langle S|
-{1\over 2}
\{\rho,|2\rangle\langle 2|\}\right)+
2\gamma^{+}_{\rm ph}
\left(\langle S| \rho |S\rangle |2\rangle\langle 2|
-{1\over 2}
\{\rho,|S\rangle\langle S| \}\right),
\end{equation}
i.e. the antisymmetric state  $|A\rangle$ is not connected by transition with the upper level (since the amplitudes of transitions between the upper level and $|1\rangle$, $|1'\rangle$ are equal the amplitudes will cancel for the antisymmetric state). Hence the state $|A\rangle$ will be a dark state.

Vibronic ''laser'' will be generated by the contribution
$$
H_{\rm eff}=s(|2\rangle\langle S|+|S\rangle\langle 2|),\quad s\in \mathbb{R}
$$
with equation of generation of the coherent mode
$$
\frac{d}{dt}s=-\varkappa s -i \rho_{2S}.
$$

We will investigate the system dynamics in the space of density matrices which contains diagonal matrix elements and one pair of off-diagonal elements (other matrix elements will decay exponentially due to decoherence)
\begin{equation}\label{rho1}
\rho=\rho_{22}|2\rangle\langle 2| + \rho_{SS}|S\rangle\langle S|+ \rho_{AA}|A\rangle\langle A| + \rho_{00}|0\rangle\langle 0| + \rho_{2S}|2\rangle\langle S|+ \rho_{S2}|S\rangle\langle 2|.
\end{equation}

\medskip

\noindent{\bf Stationary nonequilibrium state}.\quad In the stationary state $L(\rho)=0$ with $L$ given by (\ref{L1}) diagonal matrix elements of the density matrix in the case without vibrones are computed in analogy with computations of the previous section, as minors with alternating signs for the first line of the matrix $L$ in the subspace of diagonal density matrices of the form (\ref{rho1}).

In presence of vibrones off-diagonal matrix elements $\rho_{2S}$, $\rho_{S2}$ of the density matrix in the stationary state are poportional to the inversion $\rho_{22}-\rho_{SS}$ (difference of populations of the upper level and the symmetric state)
$$
\rho_{2S}=-\frac{is}{\mu}(\rho_{22}-\rho_{SS}),\qquad
\rho_{S2}=\frac{is}{\mu}(\rho_{22}-\rho_{SS}),\qquad \mu=-n_{\rm ph}\left(\gamma^{-}_{\rm ph}+\gamma^{+}_{\rm ph}\right).
$$

Therefore, as in the previous section, for diagonal elements of the stationary density matrix introduction of vibrones reduces to the transformation of transition rates of the form
$$
n_{\rm ph}\gamma^{\pm}_{\rm ph}\mapsto n_{\rm ph}\gamma^{\pm}_{\rm ph}-\frac{s^2}{\mu},
$$
and off-diagonal matrix elements are given by the above formula.

The flow of excitons to the sink equals (by expression (\ref{theta_sink1}))
$$
F=2n_{\rm sink}\left(\gamma^{-}_{\rm sink}\left(\rho_{SS}+\rho_{AA}\right)-2\gamma^{+}_{\rm sink}\rho_{00}\right)=
$$
$$
=\frac{8}{\Delta(s)}\left(4n_{\rm ex}\gamma_{\rm ex}+2n_{\rm sink}\gamma^{-}_{\rm sink}\right)
n_{\rm em}n_{\rm sink}\biggl[-\left(n_{\rm ph}\gamma^{-}_{\rm ph}- \frac{s^2}{\mu}\right)\gamma^{+}_{\rm em}\gamma^{-}_{\rm sink}
+\left(n_{\rm ph}\gamma^{+}_{\rm ph}- \frac{s^2}{\mu}\right)\gamma^{-}_{\rm em}\gamma^{+}_{\rm sink}\biggr],
$$
where $\Delta(s)$ is the normalization which makes trace of the density matrix equal to one.

This normalization has the form (let us note that $\Delta(s)$ in the present section is negative)
$$
\Delta(s)=\left(2n_{\rm ph}\gamma^{+}_{\rm ph}-\frac{2s^2}{\mu}\right)\bigl(2n_{\rm sink}\gamma^{+}_{\rm sink}\left(-2n_{\rm em}\gamma^{-}_{\rm em}+2n_{\rm sink}\gamma^{-}_{\rm sink}\right)- $$ $$
-\left(-2n_{\rm ex}\gamma_{\rm ex}-2n_{\rm sink}\gamma^{-}_{\rm sink}\right)\left(-2n_{\rm em}\gamma^{-}_{\rm em}-2n_{\rm em}\gamma^{+}_{\rm em}-4n_{\rm sink}\gamma^{+}_{\rm sink}\right)\bigr)+
$$
$$
+\left(4n_{\rm ex}\gamma_{\rm ex}+2n_{\rm sink}\gamma^{-}_{\rm sink}\right)
\bigl(\left(-2n_{\rm em}\gamma^{-}_{\rm em}-2n_{\rm ph}\gamma^{-}_{\rm ph}+ \frac{2s^2}{\mu}\right)\left(2n_{\rm sink}\gamma^{-}_{\rm sink}+2n_{\rm em}\gamma^{+}_{\rm em}+4n_{\rm sink}\gamma^{+}_{\rm sink}\right)- $$ $$ -2n_{\rm em}\gamma^{+}_{\rm em}\left(-2n_{\rm em}\gamma^{-}_{\rm em}+ 2n_{\rm sink}\gamma^{-}_{\rm sink}\right)\bigr).
$$

As in the previous section for the model of this section values of off-diagonal matrix elements and the amplitude of vibronic mode is related to the inversion $\rho_{22}-\rho_{SS}$.
In particular the condition of presence of vibronic generation is given by the existence of non-zero solution of the equation
$$
\varkappa s=-\frac{s}{\mu}(\rho_{22}-\rho_{SS}).
$$

In the stationary state the inversion has the form
$$
\rho_{22}-\rho_{SS}=\frac{1}{\Delta(s)}\biggl[\left(2n_{\rm em}\gamma^{-}_{\rm em}+2n_{\rm ph}\left(\gamma^{-}_{\rm ph}-\gamma^{+}_{\rm ph}\right)\right) $$ $$
\left(\left(2n_{\rm ex}\gamma_{\rm ex}+2n_{\rm sink}\gamma^{-}_{\rm sink}\right)\left(2n_{\rm em}\gamma^{+}_{\rm em}+4n_{\rm sink}\gamma^{+}_{\rm sink}\right)-4n_{\rm sink}^2\gamma^{+}_{\rm sink}\gamma^{-}_{\rm sink}\right)-
$$
$$
-2n_{\rm em}\gamma^{+}_{\rm em}\left(4n_{\rm ex}n_{\rm sink}\gamma_{\rm ex}\gamma^{-}_{\rm sink} +\left(2n_{\rm em}\gamma^{-}_{\rm em}+2n_{\rm sink}\gamma^{-}_{\rm sink}\right)\left(2n_{\rm ex}\gamma_{\rm ex}+2n_{\rm sink}\gamma^{-}_{\rm sink}\right)\right)\biggr].
$$

To simplify this expression let us investigate the regime where absorption of excitons in the reaction center is irreversible ($\gamma_{\rm sink}^{+}=0$) and we make the additional assumption that the rate of exchange between the symmetric and antisymmetric states $n_{\rm ex}\gamma_{\rm ex}$ is much larger than the rate of absorption of excitons in the reaction center $n_{\rm sink}\gamma^{-}_{\rm sink}$. We get (recall that in this section $\Delta(s)$ is negative)
\begin{equation}\label{inv3level2}
\rho_{22}-\rho_{SS}=-\frac{8n_{\rm em}n_{\rm ex}\gamma^{+}_{\rm em}\gamma_{\rm ex}}{\Delta(s)}\left(n_{\rm ph}\left(\gamma^{+}_{\rm ph}-\gamma^{-}_{\rm ph}\right)+2n_{\rm sink}\gamma^{-}_{\rm sink}\right).
\end{equation}

For comparison, for the model of previous section (without the degeneracy of the middle level) in the considered approximation the inversion  equals (\ref{inv3level1}), cf. below (take into account that normalizations  $\Delta(s)$ in this formula and formula (\ref{inv3level2}) above are different)
$$
\rho_{22}-\rho_{11}=\frac{
n_{\rm em}\gamma_{\rm em}^{+}\left(n_{\rm ph}\left(\gamma_{\rm ph}^{+}-\gamma_{\rm ph}^{-}\right)+n_{\rm sink}\gamma_{\rm sink}^{-}\right)}{\Delta(s)}.
$$

It is easy to see that in (\ref{inv3level2}) (where the system contains dark states $|A\rangle$)  we get the additional positive contribution for the inversion (given by the coefficient two for $n_{\rm sink}\gamma_{\rm sink}^{-}$, recall that $\gamma_{\rm ph}^{+}-\gamma_{\rm ph}^{-}<0$). This simplifies the problem to obtain positive inversion and provide vibronic generation using the scheme analogous to the described in the previous section. Therefore in the regime of small $n_{\rm sink}\gamma_{\rm sink}^{+}$ and large $n_{\rm ex}\gamma_{\rm ex}$ the effect of presence of dark states reduces to doubling of the positive contribution in the inversion, cf. formulae (\ref{inv3level1}) and (\ref{inv3level2}).

In the regime $\gamma_{\rm sink}^{+}=0$, $n_{\rm ex}\gamma_{\rm ex}>>n_{\rm sink}\gamma^{-}_{\rm sink}$ we get for the transfer rate of excitons to the sink and for the normalization the expressions
$$
F=-\frac{32}{\Delta(s)}n_{\rm ex}n_{\rm em}n_{\rm sink}\gamma_{\rm ex}\gamma^{+}_{\rm em}\gamma^{-}_{\rm sink}\left(n_{\rm ph}\gamma^{-}_{\rm ph}- \frac{s^2}{\mu}\right),
$$
$$
\Delta(s)=-8n_{\rm ex}\gamma_{\rm ex}\biggl[\left(n_{\rm ph}\gamma^{+}_{\rm ph}-\frac{s^2}{\mu}\right)
n_{\rm em}\left(\gamma^{-}_{\rm em}+\gamma^{+}_{\rm em}\right)+
\left(n_{\rm ph}\gamma^{-}_{\rm ph}- \frac{s^2}{\mu}\right)\left(2n_{\rm sink}\gamma^{-}_{\rm sink}+2n_{\rm em}\gamma^{+}_{\rm em}\right)+
$$
$$
+2n_{\rm sink}\gamma^{-}_{\rm sink} n_{\rm em}\left(\gamma^{-}_{\rm em}+\gamma^{+}_{\rm em}\right)\biggr].
$$

Analogously with the results of the previous section, {\it in presence of vibrones the transfer rate of excitons to the sink increases, but for the model with dark states the positive contribution to the inversion is twice larger compared to the model without dark states. Therefore it is easier to satisfy the condition of existence of non-zero vibronic mode}.

\section{Conclusion}

Vibronic mechanism of amplification of quantum transport of excitons is widely discussed in papers on quantum photosynthesis \cite{1203.5056}, \cite{FerrettiPCCP2013}, \cite{romero2014}, \cite{novoderezhkin2015}, \cite{srep20834}. Vibrones are collective exciton--vibrational excitations which are observed experimentally in photosynthetic systems. Unlike phonons vibrones are some modes in resonance to some transitions  for electronic states.

In the present paper we propose to describe vibronic mechanism of amplification of quantum transport by models analogous to semiclassical models of laser. In this approach vibrones are described by the coherent ''laser'' mode (i.e. by phononic laser). In presence of the coherent mode the transition rate in the system (the flow of excitons) increases.

We have considered a variation of semiclassical laser model where dissipative dynamics of quantum states of molecules is described by equations in the Lindblad form obtained by the quantum stochastic limit method. We consider two models --- a model with three level system and a model of ''laser without inversion'' with four level system (laser without inversion uses quantum dark states).

In section 2 we show that in presence of the coherent vibronic mode the rate of transport of excitons increases. Dependence of the transfer rate of excitons on  the parameters of the model is computed and the condition of generation of vibronic mode is found. Necessary condition for generation of vibronic mode is the population inversion (the upper level should have higer population than the lower).

In section 3 we consider a vibronic analog of ''laser without inversion''. The effect of inversion free laser simplifies the condition for generation of the coherent mode and start of vibronic amplification of quantum transport. For inversion free lasers the system should contain quantum ''dark'' states. Dark states in quantum photosynthesis were discussed in the literature \cite{FerrettiPCCP2013}, \cite{novoderezhkin2015}, \cite{srep20834} which supports the considered in the present paper approach.

\medskip

\noindent{\bf Acknowledgments.}\quad
This work is supported by the Russian Science Foundation under grant 17--71--20154.


\begin{thebibliography}{99}


\bibitem{Engel} Engel G.S., Calhoun T.R., Read E.L., Ahn T.-K., Mancal T., Cheng Y.-C., Blankenship R.E., Fleming G.R.
Evidence for wavelike energy transfer through quantum coherence in photosynthetic systems // Nature. 2007.
V. 446. P. 782--786.

\bibitem{Scholes} Scholes G.D., Fleming G.R., Olaya-Castro A., van Grondelle R. Lessons from nature about solar light harvesting
// Nature Chem. 2011. V. 3. P. 763--774.

\bibitem{1203.5056} A. Kolli, E.J. O'Reilly, G.D. Scholes, A. Olaya-Castro,
The fundamental role of quantized vibrations in coherent light harvesting by cryptophyte algae,
J. Chem. Phys. {\bf 137}, 174109 (2012).


\bibitem{FerrettiPCCP2013} M. Ferretti, V.I. Novoderezhkin, E. Romero, R. Augulis,
A. Pandit, D. Zigmantas, R. van Grondelle,
The nature of coherences in the B820 bacteriochlorophyll dimer revealed by two-dimensional electronic spectroscopy,
Phys. Chem. Chem. Phys. {\bf 16}, 9930--9939 (2014).

\bibitem{romero2014} E. Romero, R. Augulis, V.I. Novoderezhkin, M. Ferretti, J. Thieme, D. Zigmantas, R. van Grondelle,
Quantum coherence in photosynthesis for efficient solar-energy conversion,
Nature Physics {\bf 10}  676--682 (2014).

\bibitem{novoderezhkin2015} V.I. Novoderezhkin, E. Romero, R. van Grondelle,
How exciton-vibrational coherences control charge separation in the photosystem II reaction center,
Physical Chemistry Chemical Physics {\bf 17} 30828--30841 (2015).

\bibitem{srep20834} M. Ferretti, R. Hendrikx, E. Romero, J. Southall, R.J. Cogdell, V.I. Novoderezhkin, G.D. Scholes, R. van Grondelle,
Dark States in the Light-Harvesting complex 2 Revealed by Two-dimensional Electronic Spectroscopy,
Scientific Reports {\bf 6}:20834, (2016).

\bibitem{AcLuVo} L. Accardi, Lu Yun Gang and I. Volovich, {\it Quantum theory and its stochastic limit} (Springer-Verlag, Berlin, 2002).

\bibitem{Notes} Accardi L., Kozyrev S. Lectures on quantum interacting particle systems // Quantum interacting particle
systems. Singapore: World Scientific, 2002. P. 1--195. (QP--PQ: Quantum Probab. White Noise Anal.; V. 14).

\bibitem{Haken} H.Haken, Laser Light Dynamics, North Holland Physics Publishing, Amsterdam, New York, 1985.

\bibitem{Scully} Marlan O. Scully, M. Suhail Zubairy, Quantum Optics, Cambridge University Press, 1997.

\bibitem{Koch} Kocharovskaya O and Khanin Y I,
Coherent amplification of an ultrashort pulse in a three-level medium without population inversion,
JETP Lett. {\bf 48} 630--634 (1988).

\bibitem{Harris} Harris, S. E.  Lasers without inversion: Interference of lifetime-broadened resonances. Physical Review Letters. 62 (9): 1033--1036 (1989).

\bibitem{Mompart} Mompart, J.; Corbalan, R.  Lasing without inversion. J. Opt. B: Quantum Semiclass. Opt. 2 (3): R7--R24. (2000).

\bibitem{Dark} Fleischhauer M., Lukin M.D. Dark-state polaritons in electromagnetically induced transparency // Phys. Rev.
Lett. 2000. V. 84, N 22. P. 5094--5097; arXiv: quant-ph/0001094.

\bibitem{ArVoKo} I. Ya. Aref'eva, I. V. Volovich, S. V. Kozyrev, Stochastic limit method and interference in quantum many-particle systems, Theoretical and Mathematical Physics, 2015, 183:3, 782--799.

\bibitem{VoKo} I.V.Volovich, S.V.Kozyrev, Manipulation of States of a Degenerate Quantum System, Proceedings of the Steklov Institute of Mathematics, 2016, Vol. 294, P. 241--251.

\bibitem{KMTV} S.V. Kozyrev, A.A. Mironov, A.E. Teretenkov, I.V. Volovich,
Flows in nonequilibrium quantum systems and quantum photosynthesis, IDAQP Vol. 20, No. 4, 1750021, arXiv:1612.00213 [quant-ph] (2017)

\bibitem{Cosine} S.V. Kozyrev, Quantum transport in degenerate systems, Proceedings of the Steklov Institute of Mathematics, 2018, Vol. 301, pp. 134--143. arXiv:1709.08396

\bibitem{DarkStates} S.V. Kozyrev, I.V. Volovich, Dark states in quantum photosynthesis,
Trends in Biomathematics: Modeling, Optimization and Computational Problems, Eds. R.P.Mondaini, Springer, 2018 (Proceedings of BIOMAT 2017 Conference)

\bibitem{PechenIlyin} A. N. Pechen, N. B. Il'in, Existence of traps in the problem of maximizing quantum observable averages for a qubit at short times, Proc. Steklov Inst. Math., 289 (2015), 213--220.

\bibitem{PechenTru} Pechen A., Trushechkin A. Measurement--assisted Landau--Zener transitions // Phys. Rev. A. 2015. V. 91, N 5.
Pap. 052316.

\bibitem{OhyaVolovich} Ohya M., Volovich I. Mathematical foundations of quantum information and computation and its applications
to nano- and bio-systems. New York: Springer, 2011.

\bibitem{TruVo} Trushechkin A.S., Volovich I.V. Perturbative treatment of inter-site couplings in the local description of open
quantum networks // Europhys. Lett. 2016. V. 113, N 3. Pap. 30005.


\end{thebibliography}
\end{document}